\documentclass[manuscript,preprint]{aastex}

\slugcomment{2 November 2022 \ accepted to Astrophys.  J.}

\shorttitle{Markov Turbulence}
\shortauthors{Macek et al.}

\usepackage{lineno}
\usepackage{graphicx}
\usepackage{natbib}
\usepackage{amssymb}
\usepackage{amsmath} 
\usepackage{amsfonts}

\begin{document}

\title{\textit{Magnetospheric Multiscale} Observations of Markov Turbulence\\ 
on Kinetic Scales}


\author{Wies{\l}aw~M. Macek\altaffilmark{1,2}, Dariusz~W\'{o}jcik\altaffilmark{1,2}, 
\and James~L. Burch\altaffilmark{3}
}

\altaffiltext{1}{%
Institute of Physical Sciences, Faculty of Mathematics and Natural Sciences, 
Cardinal Stefan Wyszy\'{n}ski University, 
W\'{o}ycickiego 1/3, 01-938 Warsaw, Poland \email{macek@uksw.edu.pl}}

\altaffiltext{2}{Space Research Centre, Polish Academy of Sciences,
Bartycka 18 A, 00-716 Warsaw, Poland \email{macek@cbk.waw.pl, dwojcik@cbk.waw.pl}}

\altaffiltext{3}{Southwest Research Institute, San Antonio, TX, USA 
\email{jburch@swri.edu}}




\begin{abstract}
In our previous studies we have examined solar wind and magnetospheric plasmas turbulence, 
including Markovian character on large inertial magneto\-hydrodynamic scales. 
Here we present the results of statistical analysis of magnetic field fluctuations in the Earth's magnetosheath
based on \textit{Magnetospheric Multiscale} mission at much smaller kinetic scales.
Following our results on spectral analysis with very large slopes of about -16/3, 
we apply Markov processes approach to turbulence in this kinetic regime.
It is shown that the Chapman-Kolmogorov equation is satisfied 
and the lowest-order Kramers-Moyal coefficients describing drift and diffusion 
with a power-law dependence 
are consistent with a generalized Ornstein-Uhlenbeck process.
The solutions of the Fokker-Planck equation agree with experimental probability density functions, 
which exhibit a universal global scale invariance through the kinetic domain.
In particular, for moderate scales we have the kappa distribution 
described by various peaked shapes with heavy tails,
which with large values of kappa parameter are reduced to 
the Gaussian distribution for large inertial scales.
This shows that the turbulence cascade can be described 
by the Markov processes also on very small scales.
The obtained results on kinetic scales may be useful 
for better understanding of the physical mechanisms governing turbulence.
\end{abstract}






\section{Introduction}
\label{sec:met:int}

Turbulence appears in many real systems in nature, including various fluids with the embedded magnetic fields 
\citep{Fri95,Bis03}.
In particular, space and astrophysical plasmas are natural laboratories  
for investigating the dynamics of turbulence \citep{Cha15,BruCar16,Echet21}.
This is complex phenomenon that contains deterministic and random components.
Therefore, besides the effort to describe this problem in terms of difference equations a statistical approach is also useful. 
The important question for any dynamical system is whether given a probability distribution 
of the characteristic property of a system in a given moment, 
one can determine statistical properties of this dynamical system in a future. 
Therefore, a concept of a Markov
 process in which the future statistics 
is independent of the past is an important issue also for turbulence
\citep{PedNov94}.
It is possible to prove the existence of a Markov process experimentally and furthermore to extract 
the differential equation for this Markov process directly from the measured data 
without using any assumptions or models for the underlying stochastic process \citep{Renet01}.
\cite{StrMac08a,StrMac08b} have applied this statistical  method 
to solar wind magnetic fluctuations in the inertial range. 
A similar approach has recently been applied to the \textit{Parker Solar Probe} (PSP) mission 
in the solar wind 
at sub-proton scales \citep{Benet22}. 

Our previous studies have also dealt with turbulence 
in solar wind and magnetospheric plasmas on large-(inertial) magnetohydrodynamic scales, 
using observations by the \textit{Ulysses} mission in the solar wind beyond the ecliptic plane \citep{WawMac10},   
and \textit{Voyager} mission in the heliosphere and heliosheath \citep{Macet11,Macet12} 
and even at the boundaries of the Solar System \citep{Macet14}. 
Based on \textit{THEMIS} mission in the Earth's magnetosheath, we have also verified that 
turbulence at shocks is well described by inward and outward propagating Alfvén waves
\citep{Macet15,Macet17}.

Here we consider again turbulence in the Earth's magnetosheath, 
where timescales are much shorter than those in the heliosheath, 
but based on observations from the \textit{Magnetospheric Multiscale} (\textit{MMS}) mission
on kinetic scales \citep{Macet18}. 
In this case %
it is hotly debated whether the turbulence energy cascade results 
from the dissipation 
of the kinetic Alfv\'{e}n waves (KAW) \citep[e.g.,][]{Schet09}. 
On the contrary,
\cite{Papet21} has recently argued that the turbulence energy at kinetic scales 
could not be related to KAW activity,
but is mainly driven by localized nonlinear structures.
Certainly, it is possible that the observed stochastic nature of fluctuations in the sub-ion scale 
could be due to the interaction between coherent structures \citep[e.g.][]{Cha15,Echet21},
including local reconnection processes at kinetic scales
\citep{Macet19a,Macet19b}.
Admittedly, the nature of wave modes in operation cannot be determined 
on the statistical analysis, but we hope that the Markov approach will provide a contact point 
with a dynamical system approach to turbulence and hence  
the results of this study will be useful in future investigations.

The data under study are briefly described in Section~\ref{sec:met:d},
with statistical methods outlined in Section~\ref{sec:met:meth}. 
In Section~\ref{sec:met:res} we present the results of our analysis, 
showing that the solutions of the Fokker-Planck equation agree well  
with experimental probability density functions.
The importance of Markov processes for turbulence in space plasmas 
with a universal global scale invariance also through the kinetic domain
is underlined in Section \ref{sec:met:con}. 

\section{Data}
\label{sec:met:d}

The \textit{MMS} mission was launched in 2015 to investigate plasma processes  
in the magnetosphere and the solar wind plasma especially on small scales \citep{Buret16}.
We analyze the statistics of the fluctuations of 
all components of the magnetic field  
$\mathbf{B} = (B_x, B_y, B_z)$, with the total magnitude $B_T = |\bold B|$,
in the Geocentric Solar Ecliptic (GSE) coordinates 
obtained from the FluxGate Magnetometers (FGM), see \citep{Ruset16}.  
We investigate BURST-type observations with the highest available resolution of $\Delta t_B$ = 7.8 ms,
which corresponds to approximately 128 samples per second. 
\cite{Macet18} have selected interval on 28 December 2015 from 01.48:04 to 01.52:59  
with 37,856 measurement points for the magnetic field,
which are available just behind the bow shock (BS). 
The position of \textit{MMS} during this event within the Earth's magnetosheath 
has been depicted in Figure~1, case (a) of Ref.~\citep{Macet18}. 
Admittedly, the highest-resolution BURST-type magnetic data $B$ are limited in time. 
This analysis has allowed us to go well beyond the kinetic regime, i.e., 
above the electron Taylor-shifted inertial frequency $f_{\lambda e} = (V/c) f_{pe} $, 
where $f_{pe}$ is plasma frequency ($V$ is the solar wind velocity and $c$ denotes the speed of light),
at above $f_{\lambda e} \sim$ 20 Hz characterized by a steep spectrum with a slope of about -11/2, as seen in their Figure~2
\citep[for details see][]{Macet18}.
Even though with lower resolution for the ion velocity $V$ 
the spectrum could only be resolved to the onset of kinetic scales at $\sim$2 Hz
it is worth to investigate further this case in view of the Markov property of turbulence. 

\section{Methods}
\label{sec:met:meth}

As usual we use the increments of any characteristic parameter $x$ describing a turbulent system 
\begin{equation}
\delta x (t, \tau) = x(t+\tau) - x(t)
\label{e:met:tau}
\end{equation}
at each time $t$ and a given scale $\tau$. 
Following the well-known scenario the fluctuations $\delta x (t, \tau)$ in a larger scale are  transferred to 
smaller and smaller scales $\tau$. 
In this way turbulence may be regarded as a stochastic process with 
$N$-point joint transition probability distribution 
$P(x_1,\tau_1|x_2,\tau_2; \ldots; x_N,\tau_N)$, where
$P(x_i,\tau_i|x_j,\tau_j) = P(x_i, \tau_i; x_j, \tau_j)/P(x_j,\tau_j)$ 
is the conditional probability 
density function (PDF). 
The process is Markovian if the $N$-point joint transition probability distribution 
is completely determined by the initial values. 
Hence in this case one should have
\begin{equation}
P(x_1,\tau_1|x_2,\tau_2; \ldots; x_N,\tau_N) = P(x_1,\tau_1|x_2,\tau_2) 
\label{e:met:mar}
\end{equation}
or more generally a necessary  the Chapman-Kolmogorov condition is satisfied
\begin{equation}
P(x_1,\tau_1|x_2,\tau_2) = \int_{-\infty}^{+\infty} P(x_1,\tau_1|x',\tau') P(x',\tau'|x_2,\tau_2) d x', 
\label{e:met:chk}
\end{equation}
where $\tau_1 < \tau' < \tau_2$. 
Further, using the Kramers-Moyal expansion 
one obtains this condition in a differential form 
\begin{equation}
-\frac{\partial P(x,\tau|x',\tau')}{\partial \tau} = 
\sum_{k = 1}^{\infty}  \left(-\frac{\partial}{\partial x}\right)^{k} D^{(k)}(x,\tau)P(x, \tau|x',\tau'), 
\label{e:met:fpe}
\end{equation}
where the coefficients $D^{(k)}(x,\tau)$ are determined by the moments  
of the conditional probability density  functions \citep[cf.][]{Ris96,Benet22}  
\begin{equation} 
M^{(k)}(x,\tau, \tau') = \int_{-\infty}^{+\infty} (x' - x)^{k}  P(x',\tau'|x,\tau) d x'
\label{e:met:km} 
\end{equation}
in the limit $\tau \rightarrow \tau'$ 
\begin{equation} 
D^{(k)}(x,\tau) = \frac{1}{k!} \lim_{\tau \rightarrow \tau'} \frac{1}{\tau - \tau'} M^{(k)}(x,\tau, \tau').
\label{e:met:cef} 
\end{equation}
Moreover, if the fourth-order coefficient is equal to zero, then according to the Pawula's theorem
$D^{(k)}(x,\tau) =  0$ for $k \ge 3$, and the series is limited to the second order.  
In this case one arrives at the Fokker-Planck equation in the following reduced differential form
\citep{Ris96}:
\begin{equation}
-\frac{\partial P(x,\tau)}{\partial \tau} = 
\Big[-\frac{\partial}{\partial x} D^{(1)}(x,\tau) + \frac{\partial^{2}}{\partial x^{2}} D^{(2)}(x,\tau)\Big] P(x, \tau), 
\label{e:met:fop}
\end{equation}
where the first and second terms describe, respectively, the drift and diffusion functions 
of the deterministic evolution of the transition probability of a stochastic process 
described by the Langevin equation 
\begin{equation}
-\frac{\partial x}{\partial \tau} = D^{(1)}(x,\tau) + \sqrt{D^{(2)}(x,\tau)} \Gamma (\tau),   
\label{e:met:leq}
\end{equation} 
i.e., the process generated (with It\^{o} definition) by the delta-correlated Gaussian white noise, 
$\Gamma (\tau) \Gamma (\tau') = 2 \delta (\tau - \tau')$ \citep{Rinet16}.
The minus signs on the left-hand sides of Equations~(\ref{e:met:fop}) and (\ref{e:met:leq})
indicate that the corresponding transitions proceed backward toward smaller scales. 
More explicitly Equation~(\ref{e:met:fop}) reads  
\begin{eqnarray}
- \frac{\partial P(x,\tau)}{\partial \tau}  =    
 D^{(2)}(x,\tau) \frac{\partial^2 P(x,\tau)}{\partial x^2} 
 & + &   \Big[- D^{(1)}(x, \tau) +  2 \frac{\partial D^{(2)}(x,\tau)}{\partial x}\Big] \frac{\partial P(x,\tau)}{\partial x} + {} \nonumber\\ 
 & + &   \Big[-\frac{\partial D^{(1)}(x,\tau)}{\partial x} + \frac{\partial^{2} D^{(2)}(x,\tau)}{\partial x^{2}}\Big] P(x,\tau).
\label{e:met:full}
\end{eqnarray}
Note that here we have taken the standard definitions used by \cite{Ris96}, 
while \cite{StrMac08a,StrMac08b} and \cite{Renet01} 
have multiplied the Kramers-Moyal coefficients by $\tau$,
corresponding to a logarithmic length scale. 
A simple solution $p_s(x)$ can be obtained from the following stationary Fokker-Planck equation 
\begin{equation}
\frac{\partial }{\partial x} [D^{(2)}(x,\tau) p_s(x)] = D^{(1)}(x,\tau) p_s(x)
\label{e:met:stationary}
\end{equation}  
resulting from the left-hand side of Equation~(\ref{e:met:fop}) equal to zero. 

\section{Results}
\label{sec:met:res}

This method has been successfully applied in the inertial range for magnetic field fluctuations 
based on Ulysses data with time resolution of one second \citep{StrMac08a}.  
The Markovian character of solar wind turbulence has also been confirmed
by using ACE data for both magnetic field (16 s) and velocity (48 s) samples \citep{StrMac08b}.
In this paper we would like to test the Markov property of turbulence on much smaller millisecond scales,
which allows us to go beyond the inertial range at least for the case of magnetic field fluctuations.  

\begin{figure}[!htpb]
\centering
\includegraphics[scale=0.5]{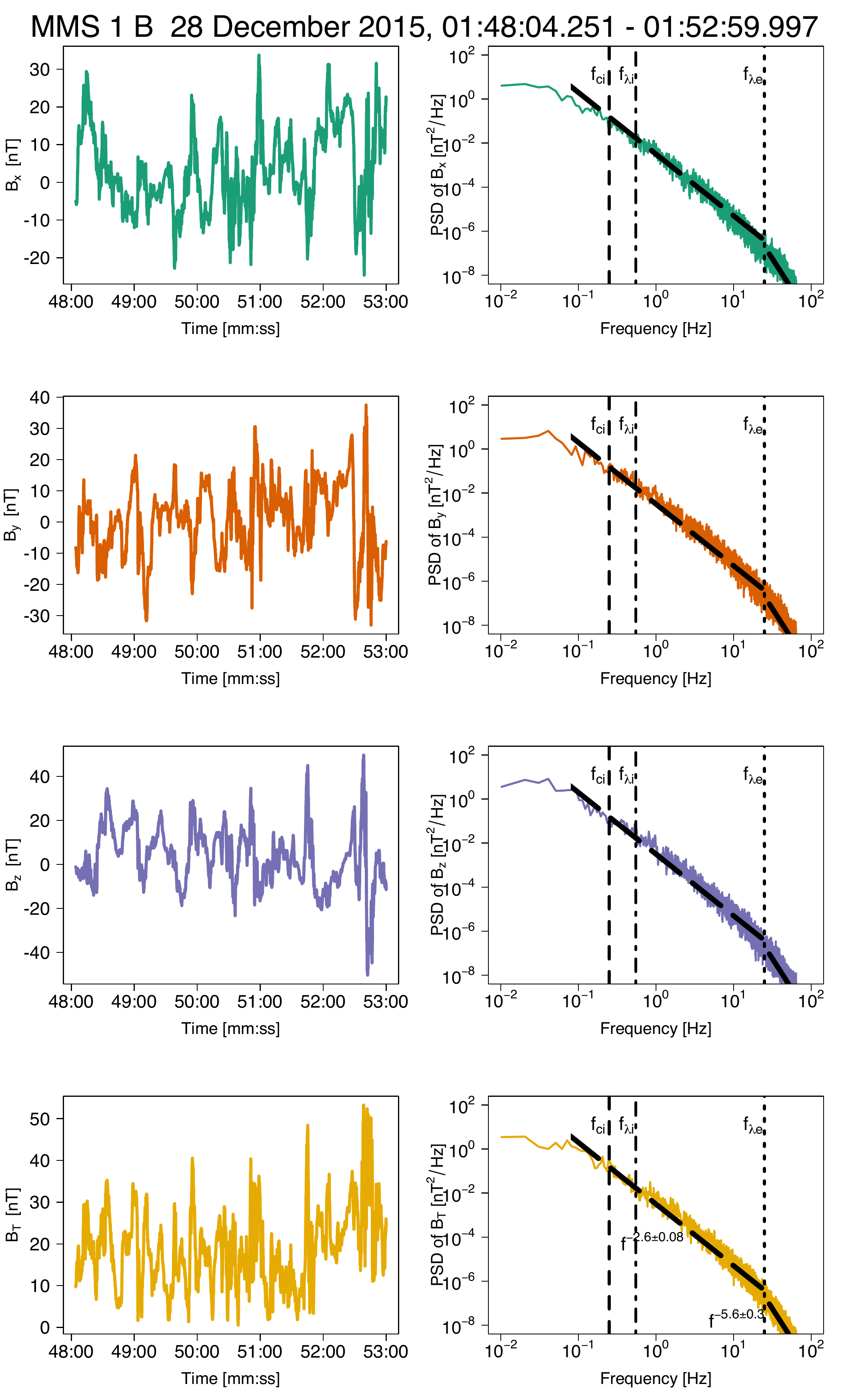}
\caption{Time series of the magnetic field (GSE) components $\bold B= (B_x,B_y,B_z)$ 
and the total magnitude $B_T = |\bold B|$ of the MMS data with the the corresponding spectrum 
of the high-resolution turbulence in the magnetosheath near the bow shock (BS),
for frequencies above the ion gyrofrequency $f_{ci}$ marked by the dashed vertical line, 
and between the ion $f_{\lambda i}$ and above the electron $f_{\lambda e}$ Taylor-shifted inertial frequencies 
shown by the dashed--dotted and dotted lines, respectively (case (a) in Table 1 of Ref. Macek et al. 2018).}
\label{f:met:bs}
\end{figure}

Figure~\ref{f:met:bs} shows 
time series of all components of the magnetic field $\bold B = (B_x,B_y,B_z)$ 
with its magnitude $B_T = |\bold B|$ in the GSM coordinates 
acquired by the MMS on 28 December 2015 during  5-minute time BURST interval (from 01.48:04 to 01.52:59),
specified as case (a) in Table 1 of Ref.~\citep{Macet18}
with  the corresponding Power Spectral Densities (PSD) 
of all the components of the magnetic field $\mathbf{B}$ 
obtained with the \citeauthor{Wel67}'s (\citeyear{Wel67}) windows. 
It is worth noting that for the magnetic spectrum above $f_{\lambda e}$ 
we enter the kinetic regime with the much steeper slope of -5.6 $\pm$ 0.3 
that is consistent with the value of -16/3  predicted by kinetic theory 
of Alfv\'{e}n waves \citep[e.g.,][]{Schet09}. 

\begin{figure}[!htpb]
\includegraphics[scale=0.55]{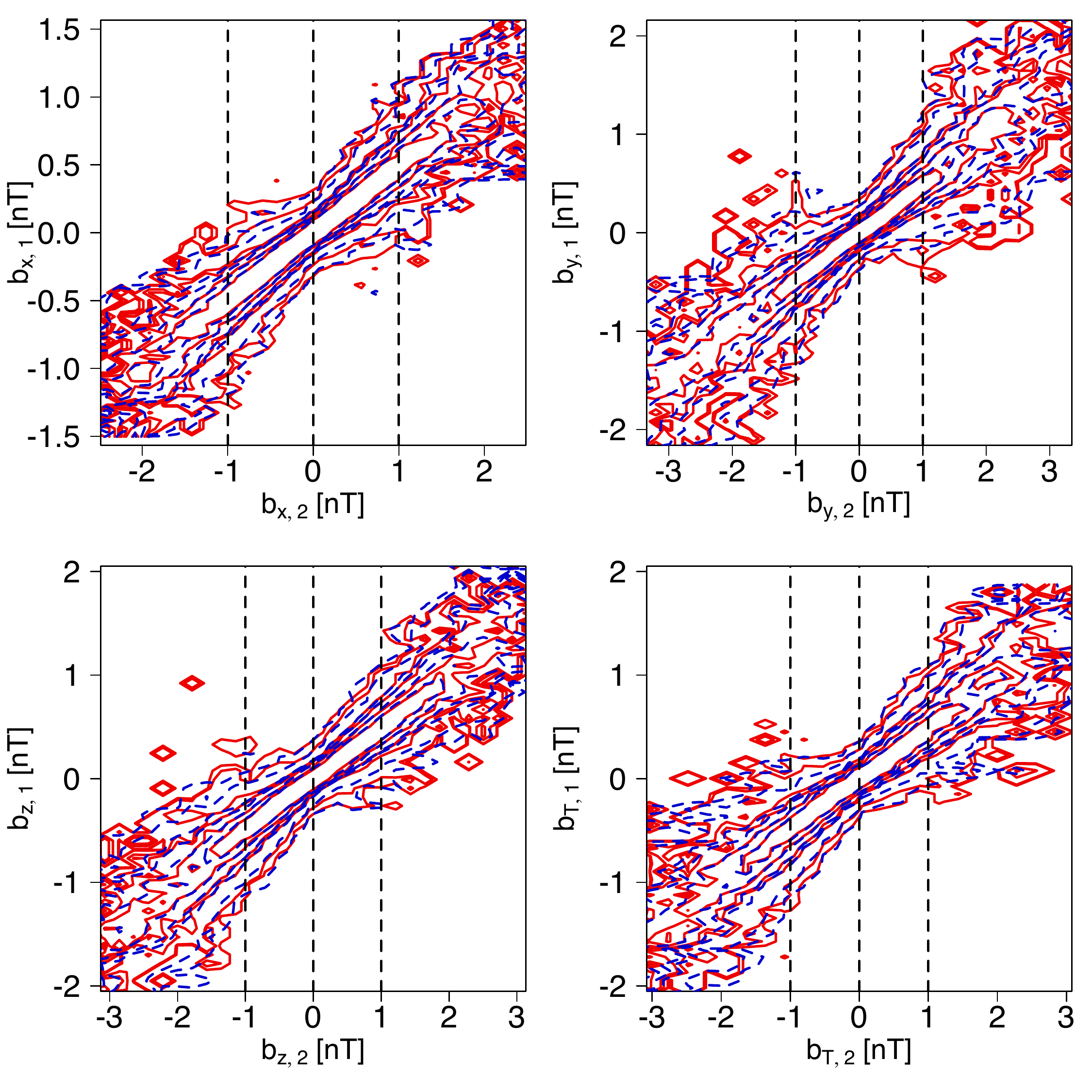}
\caption{%
Comparison of the observed (red curves) contours plots 
of conditional probabilities at various scales $\tau$
reconstructed from the MMS magnetic field components in the magnetosheath, 
corresponding to spectra in Figure~1,  
with those reconstructed (dashed blue) according to the Chapman-Kolmogorov condition, Equation~(3).
}
\label{f:met:conturs}
\end{figure}

\begin{figure}[!htpb]
\includegraphics[scale=0.55]{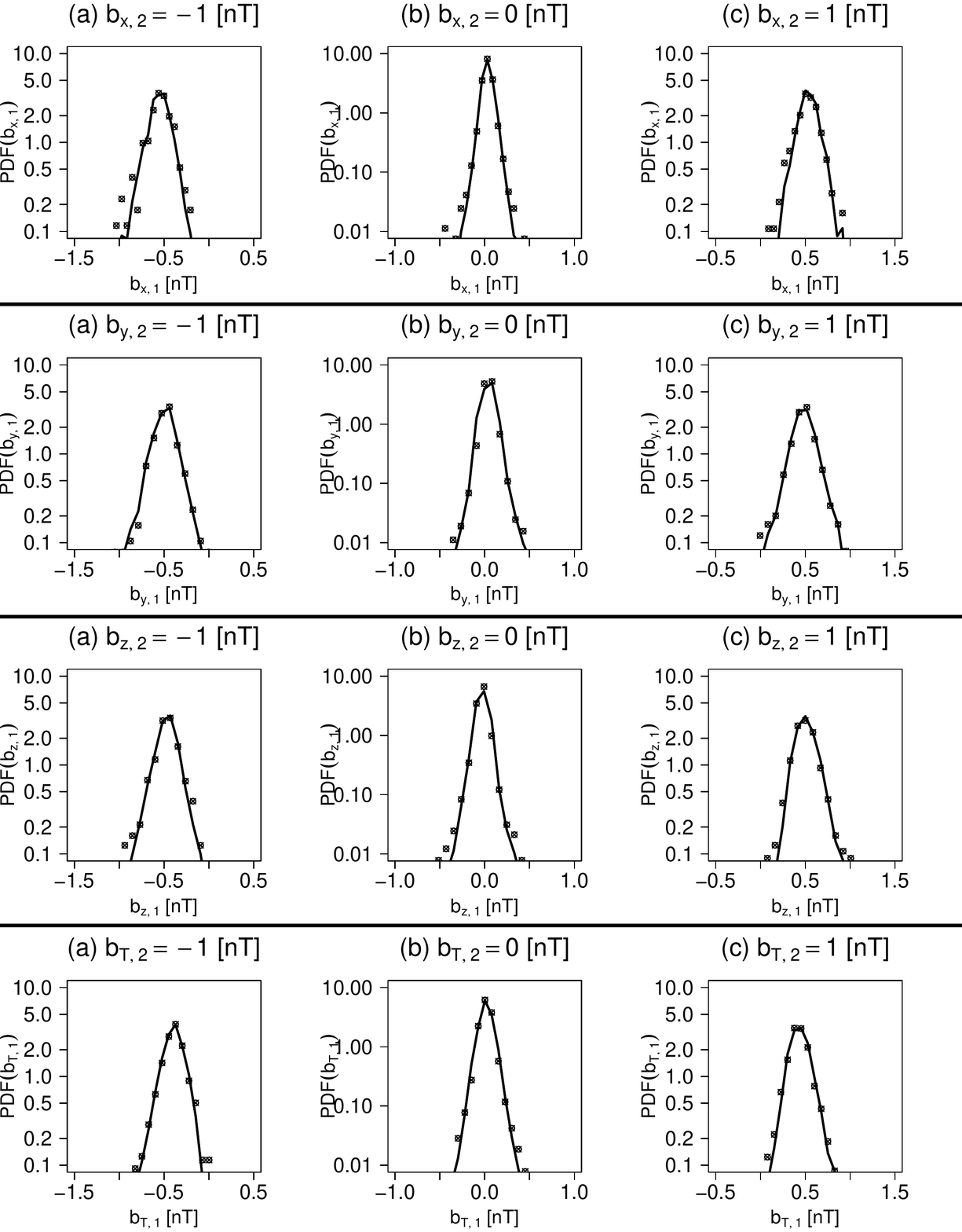}
\caption{%
Comparison of cuts through $P(\bold b_1,\tau_1|\bold b_2, \tau_2$)
for the fixed value of all components of the magnetic field increments $\bold b_2$ 
with  $\tau_1$ = 0.02 s, $\tau'$ = 0.0278 s, and $\tau_2$ = 0.0356 s.
}
\label{f:met:cuts}
\end{figure}

First, according to Equation~(\ref{e:met:tau}), we analyze 
increments of fluctuations $\bold b_\tau := \bold B(t + \tau) - \bold B(t)$
across a time scale $\tau$ 
for each GSM component $x, y, z$ and the total intensity of the magnetic field $\bold B$. 
Using the conditional probability introduced in Section~\ref{sec:met:meth},
we can compute $P(\bold b_1,\tau_1|\bold b_2,\tau_2)$ on the right hand side 
of Equation~(\ref{e:met:mar}) directly from the \textit{MMS} data. 
Then, to verify a local transfer mechanism in the turbulence cascade,  
we can  test whether the Chapman-Kolmogorov condition of Equation~(\ref{e:met:chk}) 
is satisfied for the range of scales from $\tau_1$ to $\tau_2$, 
and  $\tau_1 < \tau ’ < \tau_2$.

In Figure~\ref{f:met:conturs} we compare the observed contour plots (red curves)
of conditional probabilities at various scales $\tau$
with solutions (dashed blue) of Equation~(\ref{e:met:chk}). 
The subsequent isolines correspond to the following decreasing levels 
of the conditional probability density function (from middle of the plots)
for $\bold b$:  2, 1.1, 0.5, 0.3, 0.05, 0.02. 
In the corresponding Figure~\ref{f:met:cuts}
we verify the Chapman-Kolmogorov equation (3) by comparison of 
cuts through the conditional probability distributions 
for some chosen values of parameter $\bold b$ 
in time series specified by Equation~(\ref{e:met:tau}),
which have been differentiated and the variance stationarity has been confirmed  
by using statistical augmented Dickey - Fuller test
\citep{DicFul79}.
We have chosen here for the magnetic field $\bold b_\tau$:  
$\tau_1$ = 0.02 s, $\tau'$ = $\tau_1+ \Delta t_B$ = 0.0278 s, and $\tau_2$ = $\tau_1+ 2 \Delta t_B$  = 0.0356 s.
We see that the Equation~(\ref{e:met:chk}) is approximately satisfied up to the scales of about
100 $\Delta t_B$ = 0.78 s for $\bold b_\tau$ 
which indicates that the turbulence cascade exhibits Markov properties.

\begin{figure}[!htpb]
\includegraphics[scale=0.7]{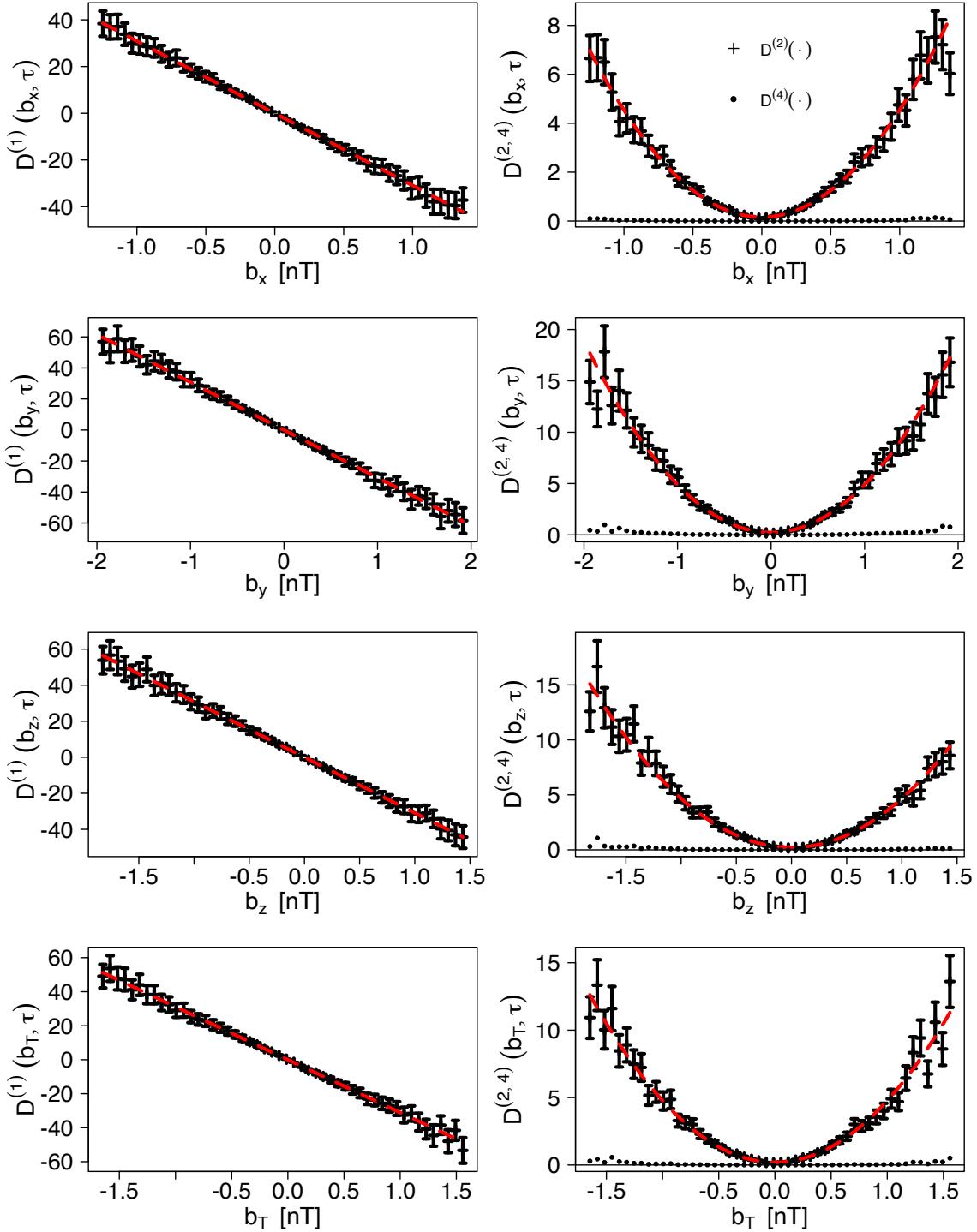}
\caption{%
First and second finite-size Kramers-Moyal coefficients 
depending on the magnetic field increments $\bold b_\tau$
for all components of the magnetic field $\mathbf{B} = (B_x, B_y, B_z)$ and the total magnitude $B_T = |\bold B|$.
The dashed red lines show the best fits to the calculated values of $D^{(1)}(\bold b,\tau)$ and $D^{(2)}(\bold b,\tau)$
with $D^{(4)}(\bold b,\tau)$ = 0, according to the Pawula's theorem.
}
\label{f:met:km}
\end{figure}

Second, we need to compute the Kramers-Moyal coefficients $D^{(k)}(x,\tau)$ 
in the Fokker-Planck expansion given by Equation~(\ref{e:met:fpe}).
The values of the moments $M^{(k)}(x,\tau, \tau')$ defined in Equation~(\ref{e:met:km}) 
can be obtained from the experimental data 
by counting the number $N(x', x)$ of occurrences of fluctuations $x'$ and $x$. 
Since the errors of $N(x ', x)$ are given by  $1/\sqrt{N(x ', x)}$ 
the errors for the conditional moments $M^{(k)} (x, \tau, \tau ')$ 
can also be provided \citep[see,][]{Renet01}. 

Admittedly $D^{(k)}(x,\tau)$ can only be obtained by extrapolation 
(for instance using piecewise linear regression) 
in the limit $\tau  \rightarrow \tau'$ in Equation~(\ref{e:met:cef}), but  
we have checked that the very similar values are obtained taking 
the lowest time resolution $\tau-\tau' = \Delta t_B$ = 0.0078 s. 
In fact, we have 
$D^{(k)}(x,\tau) \approx \frac{1}{k!} \frac{1}{\Delta t} M^{(k)}(x,\tau, \tau')$.
Therefore, basically the coefficients $D^{(k)} (x, \tau)$ show the same dependence on $x$ as $M^{(k)} (x, \tau, \tau‘)$ 
\citep[cf.][]{Renet01}. 
Figure~\ref{f:met:km} presents the fits to the first order drift $D^{(1)}(x,\tau)$ 
and the second order finite-size diffusion  $D^{(2)}(x,\tau)$ coefficients
for $\Delta t_B$ = 0.0078 s. 
We have also verified that the fourth-order coefficient $D^{(4)}(x,\tau)$ is close to zero for   
magnetic field data according to the Pawula's theorem, 
which is a necessary and sufficient condition that the Kramers-Moyal expansion 
of Equation~(\ref{e:met:fpe}) stops after the second term.

In this case, we see that the best obtained fits to these lowest order coefficients are linear 
\begin{equation}
D^{(1)}(x,\tau) = -a_1(\tau) x  
\label{e:met:fit1}
\end{equation}
and quadratic functions of $x$
\begin{equation}
D^{(2)}(x,\tau) = a_2(\tau) + b_2(\tau) x^2, 
\label{e:met:fit2}
\end{equation}
respectively, where the appropriate fitted parameters 
$a_k$ for $k$ = 1 and 2 and $b_2$ depend on the time scale $\tau$.
This corresponds to the generalized Ornstein-Uhlenbeck process. 
It is interesting to note that, similarly as obtained for the \textit{PSP} data by \cite{Benet22}, 
who have looked at larger $\tau$, 
the best fit for any $x$ representing each component of $\bold b_\tau$ 
satisfies a power-law dependence $A \tau^{\alpha}$ with sufficient accuracy 
and the values for all the parameters are listed in Table~1.

\begin{figure}[!htpb]
\includegraphics[scale=0.7]{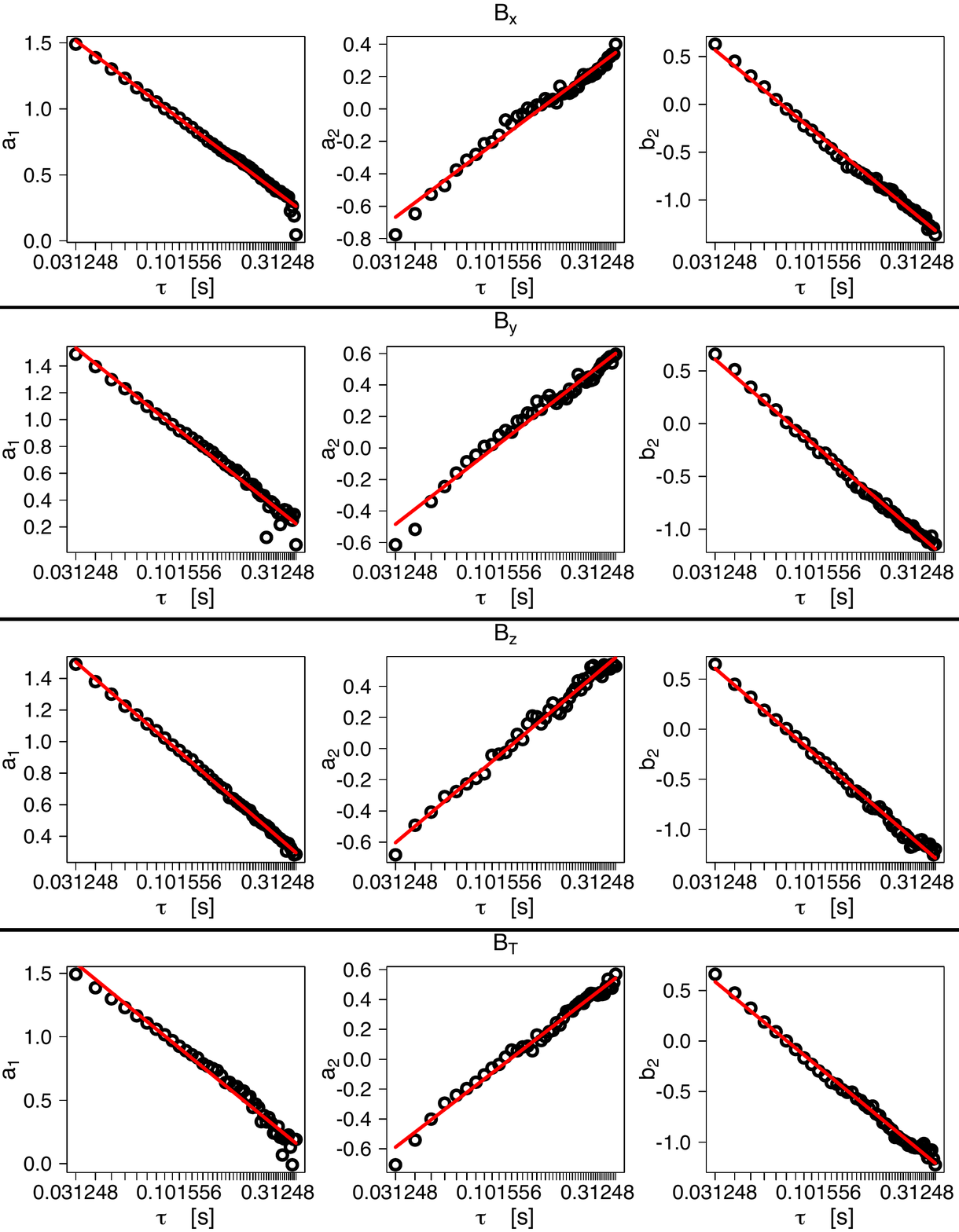}
\caption{%
Linear dependence of the parameters $a_1$, $a_2$, and $b_2$ on logarithmic scale $\tau$, 
see Equations~(11) and (12), 
for all components of the magnetic field $\mathbf{B} = (B_x, B_y, B_z)$ and the total magnitude $B_T = |\bold B|$
} 
\label{f:met:fit}
\end{figure}

\begin{table}[htbp]
\caption{Parameters for Power-law Dependence of First and Second Kramers-Moyal Coefficients
corresponding to Equations~(11) and (12)
for all components of the magnetic field $\mathbf{B}$ and the total magnitude $B_T = |\bold B|$
}
\parbox{.41\linewidth}{
\centering
\begin{tabular}{|r|r|r|}
\hline
\boldmath$b_x$ & \boldmath$A$ & \boldmath$\alpha$ \\
\hline
\hline
\boldmath$a_1$ & 0.6638 $\pm$ 0.0355 & -1.2376 $\pm$ 0.0215 \\
\hline
\boldmath$a_2$ & -0.4925 $\pm$ 0.0155 & 0.9416 $\pm$ 0.0094 \\
\hline
\boldmath$b_2$ & 0.5918 $\pm$ 0.0296 & -1.6919 $\pm$ 0.0179 \\
\hline
\end{tabular}
}
\hfill
\parbox{.41\linewidth}{
\centering
\begin{tabular}{|r|r|r|}
\hline
\boldmath$b_y$ & \boldmath$A$ & \boldmath$\alpha$ \\
\hline
\hline
\boldmath$a_1$ & 0.6534 $\pm$ 0.0278 & -1.2026 $\pm$ 0.0169 \\
\hline
\boldmath$a_2$ & -0.4216 $\pm$ 0.0203 & 0.9699 $\pm$ 0.0123 \\
\hline
\boldmath$b_2$ & 0.5612 $\pm$ 0.0316 & -1.5263 $\pm$ 0.0192 \\
\hline
\end{tabular}
}
$\newline \newline \\$
\parbox{.41\linewidth}{
\centering
\begin{tabular}{|r|r|r|}
\hline
\boldmath$b_z$ & \boldmath$A$ & \boldmath$\alpha$ \\
\hline
\hline
\boldmath$a_1$ & 0.5646 $\pm$ 0.0286 & -1.2253 $\pm$ 0.0173 \\
\hline
\boldmath$a_2$ & -0.4024 $\pm$ 0.0172 & 1.0934 $\pm$ 0.0104 \\
\hline
\boldmath$b_2$ & 0.5941 $\pm$ 0.0241 & -1.6623 $\pm$ 0.0146 \\
\hline
\end{tabular}
}
\hfill
\parbox{.42\linewidth}{
\centering
\begin{tabular}{|r|r|r|}
\hline
\boldmath$b_T$ & \boldmath$A$ & \boldmath$\alpha$ \\
\hline
\hline
\boldmath$a_1$ & 0.6989 $\pm$ 0.0225 & -1.1191 $\pm$ 0.0089 \\
\hline
\boldmath$a_2$ & -0.4946 $\pm$ 0.1259 & 1.1631 $\pm$ 0.0498 \\
\hline
\boldmath$b_2$ & 0.5854 $\pm$ 0.0706 & -1.7325 $\pm$ 0.0279 \\
\hline
\end{tabular}
}
\end{table}

In fact, as seen in Figure~\ref{f:met:fit} on logarithmic scale \citep[cf.][Fig.~5]{StrMac08a}, 
we have verified here that for the \textit{MMS} magnetic field data 
these lowest-order fits with power-law dependence 
apply for $\tau \le 100 \Delta t_B = \ 0.78$ s,
when the probability density function is closer to Gaussian, $\tau \sim \tau_{G}$. 
However, for higher scales more complex functional dependence is necessary, 
especially in the inertial regime \citep{Renet01,StrMac08a,StrMac08b} 

Anyway, we see again that using the simple linear and parabolic fits of Equations~(\ref{e:met:fit1}) and (\ref{e:met:fit2}),
the stationary solutions of Equation~(\ref{e:met:stationary})
become the well-known continuous kappa distributions 
(also known as Pearson type VII distribution), 
which probability density function is defined as:
\begin{equation}
p_s(x) = \frac{N_o}{\Big[1 + \frac{1}{\kappa} \Big(\frac{x}{x_o}\Big)^2\Big]^{\kappa}}
\      = \frac{N'_o}{\Big[a_2(\tau) + b_2(\tau) x^2\Big]^{1 + \frac{a_1(\tau)}{2 b_2(\tau)}}} 
\label{e:met:kappa}
\end{equation}  
with $\kappa = 1 + a_1(\tau)/[2 b_2(\tau)]$ and $x_o^2 = a_2 (\tau) / b_2(\tau ) / \kappa = a_2 (\tau) /[b_2 (\tau) + a_1(\tau)/ 2]$  
(for $a_2(\tau) \ne 0$, $x_o(\tau) \ne 0$)
and $N_o = p_s(0)$ satisfying  the normalization $\int_{-\infty}^{+\infty} p_s(x') dx ' = 1$, i.e.,
$N_o = \frac{\Gamma(\kappa)}{x_o \sqrt{\pi \kappa} \Gamma(\kappa -1/2)}$.
As requested, the boundary condition $p_S (x \rightarrow \pm \infty) \rightarrow 0$ is also verified here, 
and with $\kappa \rightarrow \infty$ the distribution degenerates 
into the Maxwellian distribution $N_o \mathrm{e}^{-(\frac{x}{x_o})^2}$
with $N_o = \frac{1}{x_o \sqrt{\pi}}$.
The values of the relevant parameters of Equation~(\ref{e:met:kappa}) obtained by fitting 
the \textit{MMS} data the given distributions are:
$\kappa$ = 11.85673, $x_0$ = 1.756009, $N'_0$ = 0.4121234, for $B_x$;
$\kappa$ = 10.09043, $x_0$ = 3.05319, $N'_0$ = 0.2684886, for $B_y$;
$\kappa$ = 11.04104, $x_0$ = 2.779299, $N'_0$ = 0.2802258, for $B_z$;
$\kappa$ = 12.88198, $x_0$ = 1.75533, $N'_0$ = 0.4133008, for $B_T$.
These values of $\kappa$ would correspond to the nonextensivity parameter 
of the generalized (Tsallis') entropy $q = 1 - 1/\kappa \approx 0.9$, 
which is somewhat larger than $q \sim 0.5$ for $\kappa \sim 2$
reported for the \textit{PSP} data by \cite{Benet22}. 

In addition, substituting Equations~(\ref{e:met:fit1}) and (\ref{e:met:fit2}) 
into Equation~(\ref{e:met:fop}) we obtain 
\begin{equation}
[a_2(\tau) + b_2(\tau) x^2] \frac{\partial^2 P(x,\tau)}{\partial x^2} 
 + [a_1(\tau) + 4 b_2 (\tau)] x \frac{\partial P(x,\tau)}{\partial x} 
 + \frac{\partial P(x,\tau)}{\partial \tau}
 + [a_1(\tau) + 2 b_2 (\tau)] P(x, \tau) = 0.
\label{e:met:parabolic}
\end{equation}
This means that in the Fokker-Planck Equations~(\ref{e:met:full}) and (\ref{e:met:parabolic}) 
become formally the second order parabolic partial differential equation. 

\begin{figure}[!htpb]
\includegraphics[scale=0.6]{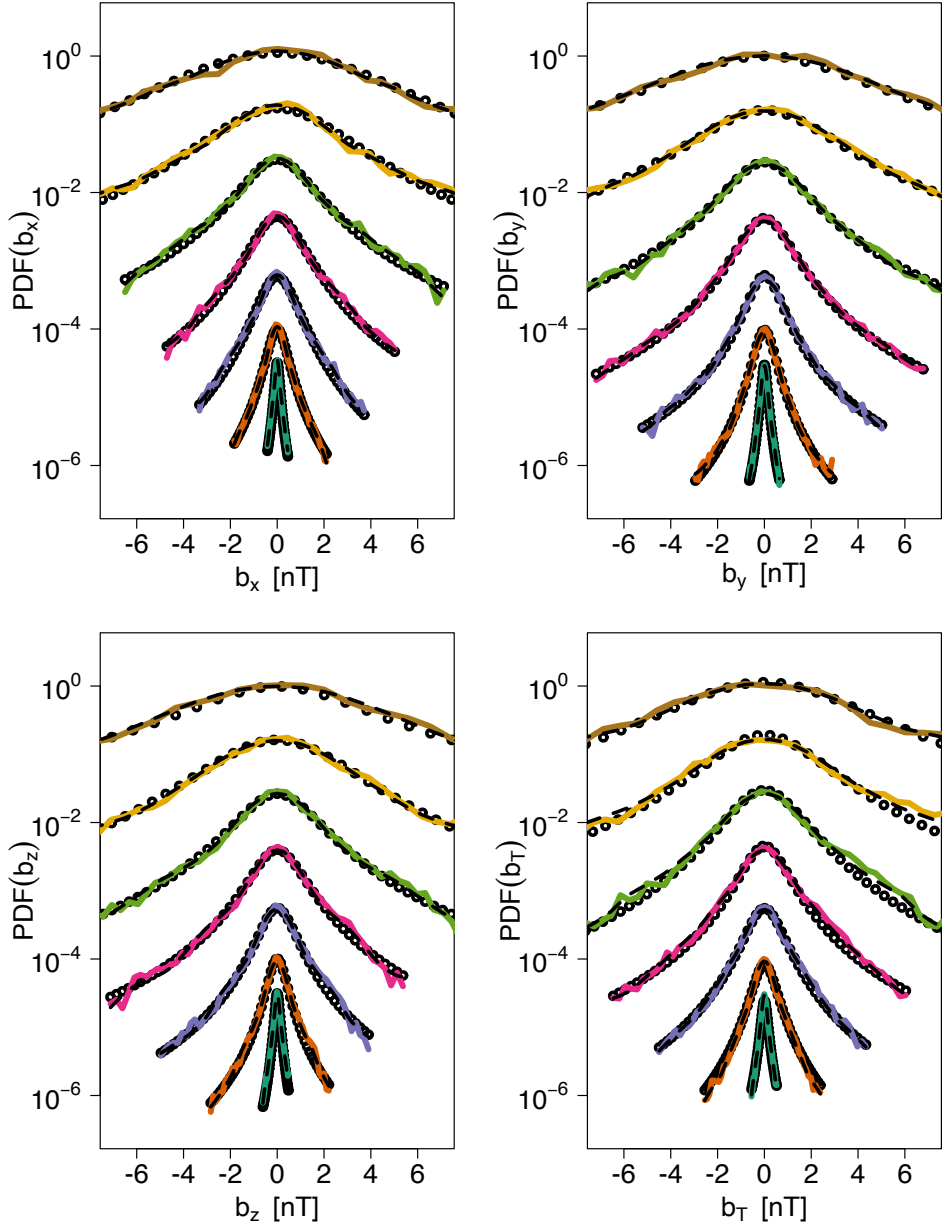}
\caption{%
Comparison of the non-stationary (dashed lines) and the stationary (open points) solutions of the Fokker-Planck equation 
with the experimental probability density functions (PDF) of the magnetic field components $P(\bold b,\tau)$  
fluctuations (continuous colored lines) 
for all components of the magnetic field $\mathbf{B} = (B_x, B_y, B_z)$ and the total magnitude $B_T = |\bold B|$
in the magnetosheath behind the bow shock using the MMS data, 
corresponding to  spectra in Fig.~1,
for various scales (shifted from bottom to top) $\tau$ = 0.0078, 0.04, 0.078, 0.12, 0.2, 0.39, and 0.78 s. 
}
\label{f:met:fokker}
\end{figure}

\begin{figure}[!htpb]
\includegraphics[scale=0.7]{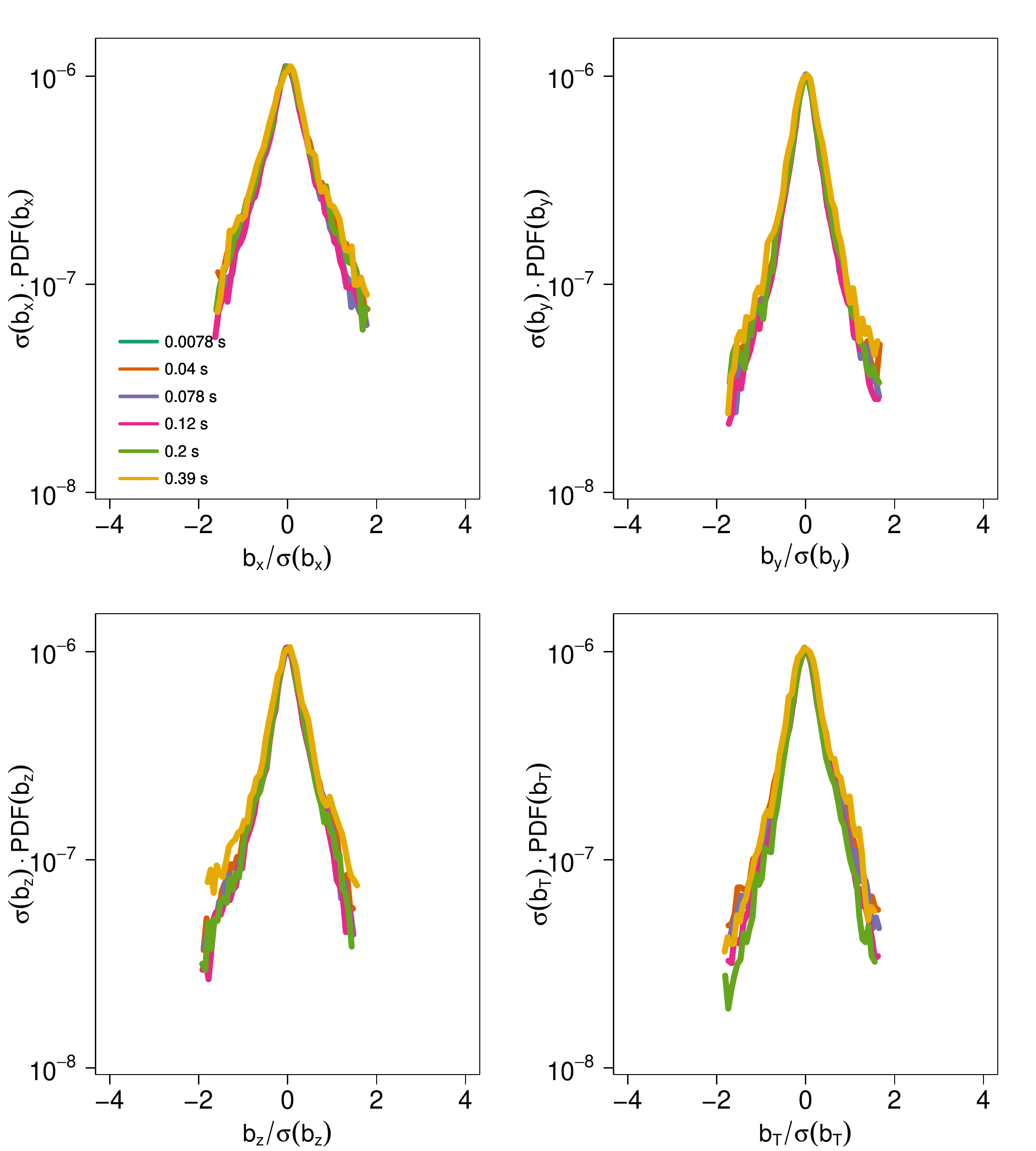}
\caption{%
A universal scale invariance of the collapsing probability density functions of $\bold b_\tau$
rescaled by the respective standard deviations $\sigma_{\bold b, \tau}$,
corresponding to the kappa distributions in Figure 6 on the kinetic scales up to $\tau \sim 0.4$ s.
}
\label{f:met:collaps}
\end{figure}

This in turn allows us to solve numerically non-stationary Fokker-Planck equation (dashed lines) using the numerical 
Euler integration scheme (verified for stationary solution $\frac{\partial P(x,\tau)}{\partial \tau} = 0$,
open points), which agrees with those obtained with modeling package by \cite{Rinet16}. 
We compare all these theoretical solutions with 
the probability density functions obtained directly from experimental data 
denoted by different colored continuous lines.
This comparison is depicted in Figure~\ref{f:met:fokker} 
for various scales $\tau$, not greater than $\tau_{G}$,
namely (from bottom to top) for $\tau$ = 1, 5, 10, 15, 25, 50, and 100 $\Delta t_B$  
shifted in the vertical direction for clarity of presentation. 
For moderate values up to $\tau \sim 50 \Delta t_B$ = 0.39 s 
in the case of linear and parabolic fits to Equations~(\ref{e:met:fit1}) and (\ref{e:met:fit2}),
we have the kappa distributions. 
However if we move to the larger scales $\tau_{G}$ from  100 $\Delta t_B$ = 0.78 s,
the Kramers-Moyal coefficients $D^{(1)}$ and $D^{(2)}$ in Equation~(\ref{e:met:fpe})
are possibly described by more complex polynomial functions, 
but the probability density function is approximately Gaussian,
as is expected for large values of $\kappa$.
On the other hand, for the smallest available scales 
we see a very peaked density function (with large kurtosis),
well described by the approximate shape of the Dirac delta function
(formally in the limit of $\tau \rightarrow 0$).

In Figure~\ref{f:met:collaps}, we have finally reproduced probability density functions of all components 
$\bold b_\tau$ rescaled by the respective standard deviations $\sigma_{\bold b, \tau}$,
which are consistent with the stationary solutions (open points) in Figure~\ref{f:met:fokker}.
Owing to a power-law dependence of the the first and second  Kramers-Moyal parametrisation,
as for \textit{PSP} analysis by \cite{Benet22}, 
near the Sun
with $\kappa \sim2$ for scales up to $\tau \sim0.05$ s, 
\textit{MMS} data exhibit a universal global scale invariance
mainly at 1 AU up to
$\tau \sim 50 \Delta t_B \sim0.4$ s, 
where we have clear kappa distributions, but with some larger values of $\kappa \sim 10$. 
Since for somewhat larger scales from $\tau_G \sim 100 \Delta t_B \sim0.8$ s (not shown here) 
the respective kappa distributions are very close to a limiting Gaussian shape, 
this would result in some more deviations from the global scale invariance (not only on tails). 

Our results 
demonstrate that the energy transfer among the different scales 
is essentially a stochastic process that can be modelled by 
the Fokker-Planck advection-diffusion equation also in the kinetic regime.
As we already suggested in our earlier analysis for inertial range of scales \citep{StrMac08a,StrMac08b}, 
because the transfer among the different scales is a stochastic ‘memoryless’ process, 
we should expect a universal structure in the turbulent dynamics. 
This is actually shown by our statistical analysis of the probability density functions
up to kinetic scales. 

\section{Conclusions}
\label{sec:met:con}

\textit{Magnetospheric Multiscale} and \textit{Parker Solar Probe} missions 
with unprecedented high millisecond time resolution
of magnetometers data allow us 
to investigate turbulence on very small kinetic scales.
In this paper we have looked at the \textit{MMS} observations above 20 Hz, 
where the magnetic spectrum becomes very steep 
with the slope close to -16/3 resulting possibly from  
interaction between coherent structures.

Following our previous studies in the inertial region \citep{StrMac08a,StrMac08b}
we have shown for the first time that the Chapman-Kolmogorov equation,
which is a necessary condition for Markovian character of turbulence, 
is satisfied exhibiting a local transfer mechanism of turbulence cascade 
also on much smaller kinetic  scales.
Moreover, we have verified that in this case 
the Fokker-Planck equation is reduced to drift and diffusion terms
at least for scales smaller than 0.8 s.

In particular, similarly as for Parker Solar probe data analyzed by \cite{Benet22} 
these lowest order coefficients are linear and quadratic functions of magnetic field,
which correspond to the generalized Ornstein-Uhlenbeck processes.
We have also recovered a similar universal scale invariance 
of the probability density functions up to kinetic scales of about 0.4 s.

It is interesting to note that for moderate scales 
we have also non-Gaussian (kappa) distribution,
which  for the smallest values of the available scale of 7.8 ms, 
is approximately described by a very peaked shape close to Dirac delta function.
We also show that the normal Gaussian distribution is recovered 
for timescale two order larger (with large value of kappa parameter).

We hope that our observation of Markovian futures in solar wind turbulence 
could be important for understanding the relation between deterministic and stochastic properties 
of turbulence cascade at kinetic scales in complex astrophysical systems. 

%
%

\acknowledgments
We thank Marek Strumik for discussion on the theory of Markov processes. 
We are grateful for the dedicated efforts of the entire \textit{MMS} mission team, 
including development, science operations, 
and the Science Data Center at the University of Colorado.
We especially benefited from the efforts of  T.~E. Moore as a Project Scientist, 
C.~T. Russell and the magnetometer team for providing the magnetic field data
for providing the data that are available online from \textrm{http://cdaweb.gsfc.nasa.gov}.
We acknowledge B.~L. Giles, Project Scientist for information about the magnetic field instrument,
and also to D. G. Sibeck and M.~V.~D. Silveira for discussions 
during previous visits by W.M.M to the NASA Goddard Space Flight Center.
We would like to thank the referee for inspiring comments,
especially on a universal scale invariance through the kinetic domain.
This work has been supported by the National Science Centre, Poland (NCN), through grant No. 2021/41/B/ST10/00823. 

ORCID iDs

\noindent
W. M. Macek https://orcid.org/0000-0002-8190-4620\\
D. W\'{o}jcik https://orcid.org/0000-0002-2658-6068\\
J. L. Burch https://orcid.org/0000-0003-0452-8403

\hyphenation{Post-Script Sprin-ger}

\end{document}